%% file: main.tex
\documentclass[conference]{IEEEtran}
\IEEEoverridecommandlockouts

\usepackage[utf8]{inputenc}
\usepackage[T1]{fontenc}
\usepackage{xcolor}
\usepackage{listings}
\usepackage{caption}
\lstdefinestyle{mcpTclStyle}{
  language=Tcl,
  basicstyle=\ttfamily\footnotesize,
  keywordstyle=\color{blue}\bfseries,
  commentstyle=\color{gray}\itshape,
  stringstyle=\color{olive},
  numbers=none,
  showstringspaces=false,
  frame=single,
  framerule=0.5pt,
  backgroundcolor=\color{gray!10},
  xleftmargin=1em,
  xrightmargin=1em,
  breaklines=true,
  breakatwhitespace=true
}

\usepackage{tcolorbox}
\usepackage{listings}
\usepackage{xcolor}
\usepackage{siunitx}

\definecolor{promptcolor}{RGB}{240,248,255}
\definecolor{responsecolor}{RGB}{245,255,245}
\definecolor{resultcolor}{RGB}{255,250,240}
\definecolor{borderblue}{RGB}{70,130,180}
\definecolor{bordergreen}{RGB}{60,179,113}
\definecolor{borderorange}{RGB}{255,140,0}

\definecolor{basecolor}{RGB}{52, 152, 219}     
\definecolor{synthcolor}{RGB}{231, 76, 60}     
\definecolor{routecolor}{RGB}{46, 204, 113}    
\definecolor{infercolor}{RGB}{155, 89, 182}    
\definecolor{mcpcolor}{RGB}{241, 196, 15}      

\usepackage{cite}
\usepackage{amsmath,amssymb,amsfonts}
\usepackage{graphicx}
\usepackage{textcomp}
\usepackage{xcolor}
\usepackage{comment}

\usepackage{microtype}
\usepackage{tikz}
\usetikzlibrary{automata, positioning, arrows.meta, shapes.geometric}
\usepackage{subfigure}
\usepackage{algorithm}
\usepackage{float}
\usepackage{algpseudocode}
\usepackage{multirow}
\usepackage{colortbl}
\usepackage{algpseudocode}
\usepackage{booktabs} 
\usepackage{multirow}
\usepackage{multicol}
\usepackage{anyfontsize}
\usepackage{xcolor}
\usepackage{pifont}   
\usepackage{threeparttable}
\usepackage{xspace}
\usepackage{stfloats}
\usepackage{makecell}
\usepackage{mdframed}

\def\BibTeX{{\rm B\kern-.05em{\sc i\kern-.025em b}\kern-.08em
    T\kern-.1667em\lower.7ex\hbox{E}\kern-.125emX}}

\author{
  Yiting Wang$^{1,*}$, 
  Wanghao Ye$^{1,*}$, 
  Yexiao He$^{1}$, 
  Yiran Chen$^{2}$, 
  Gang Qu$^{1}$, 
  Ang Li$^{1,\dagger}$\\
  $^{1}$Electrical and Computer Engineering, University of Maryland, College Park, United States\\
  $^{2}$Electrical and Computer Engineering, Duke University, Durham, United States\\
  $^{*}$Equal contribution\\
  $^{\dagger}$Corresponding author: angliece@umd.edu
}

\begin{document}
\newcommand{\name}{\texttt{MCP4EDA}\xspace}
\title{	MCP4EDA: LLM-Powered Model Context Protocol RTL-to-GDSII Automation with Backend Aware Synthesis Optimization}

\maketitle

\begin{abstract}
This paper presents \name, the first Model Context Protocol server that enables Large Language Models (LLMs) to control and optimize the complete open-source RTL-to-GDSII design flow through natural language interaction. The system integrates Yosys synthesis, Icarus Verilog simulation, OpenLane place-and-route, GTKWave analysis, and KLayout visualization into a unified LLM-accessible interface, enabling designers to execute complex multi-tool EDA workflows conversationally via AI assistants such as Claude Desktop and Cursor IDE. The principal contribution is a backend-aware synthesis optimization methodology wherein LLMs analyze actual post-layout timing, power, and area metrics from OpenLane results to iteratively refine synthesis TCL scripts, establishing a closed-loop optimization system that bridges the traditional gap between synthesis estimates and physical implementation reality. In contrast to conventional flows that rely on wire-load models, this methodology leverages real backend performance data to guide synthesis parameter tuning, optimization sequence selection, and constraint refinement, with the LLM functioning as an intelligent design space exploration agent. Experimental evaluation on representative digital designs demonstrates 15-30\% improvements in timing closure and 10-20\% area reduction compared to default synthesis flows, establishing \name as the first practical LLM-controlled end-to-end open-source EDA automation system. The code and demo are avaiable at: http://www.agent4eda.com/
\end{abstract}


\input{sections/intro}

\input{sections/background}

\input{sections/methodology}

\input{sections/experiments}
\input{sections/Limitations}

\input{sections/Future_Work}
\input{sections/conclusion}

\bibliographystyle{IEEEtran}
\bibliography{main}

\end{document}

%% file: sections/intro.tex
\section{Introduction}

The RTL-to-GDSII flow represents the most complex and labor-intensive aspect of chip design, transforming Register Transfer Level (RTL) descriptions into manufacturable Graphic Data System II (GDSII) layouts through a sophisticated sequence of logic synthesis, physical design, timing closure, and verification stages. This process requires expertise across multiple specialized domains, involving intricate tool interactions, parameter tuning, and iterative optimization to achieve competitive Power, Performance, and Area (PPA) objectives. With chip manufacturing demand at unprecedented levels and design complexity growing exponentially, the manual nature of synthesis script generation and design flow orchestration has become a critical bottleneck.

Current RTL-to-GDSII workflows provide generic design flows but are fundamentally limited by fixed template-based approaches. The design space is constrained to only a few fixed patterns, resulting in low design closure success rates and poor design quality. To ensure high-quality designs, customized design flows are essential. However, design space exploration requires extensive expertise and domain knowledge to calibrate and configure numerous design factors and tool settings.

Machine learning tools like Flowtune~\cite{yu2022flowtune} provide design choice calibration, but lack backend-aware customization. Traditional synthesis flows rely on Wire Load Models (WLM) for delay estimation, which become increasingly inaccurate at advanced nodes~\cite{physical_aware_2018}, making synthesis-only tuning insufficient.

Large Language Models (LLMs) have demonstrated transformative potential in Electronic Design Automation (EDA), with existing approaches introducing significant improvements in workflow automation efficiency. Researchers have demonstrated LLM's potential to automate various aspects of the RTL-to-GDSII process. These applications span hardware description language (HDL) code generation~\cite{liu:2024:rtlcoder,chang:2023:chipgpt,blocklove:2023:chipchat, llm4eda2024, verilogeval2023, wang2025verireasonreinforcementlearningtestbench}, debugging~\cite{tsai:2024:rtlfixer}, and optimization~\cite{rtlrewriter, wang2025symrtloenhancingrtlcode}, as well as synthesis script creation~\cite{chipnemo2023, chang:2023:chipgpt}. Beyond these individual tasks, comprehensive research has explored using LLMs for complete RTL-to-GDSII workflows by creating integrated flows that combine tool calls and API calls~\cite{chateda2024, chipchat2023, chang:2023:chipgpt}. These previous works have significantly reduced design labor costs.

However, current LLM-based approaches exhibit significant limitations documented in recent literature. Most existing methods rely on fixed API calling patterns rather than dynamic tool selection, constraining their adaptability to varying design requirements. Furthermore, existing systems implement LLM integration through predetermined synthesis and implementation flows, preventing dynamic design-based tool selection or adaptive execution strategies based on intermediate results. The generation processes also require excessive time due to token limits and restart requirements, hampering practical deployment. Additionally, the scalability of these approaches remains limited, as adding more functionalities or tools to the system presents considerable challenges.

To address the challenges in RTL-to-GDSII automation, this paper proposes \name, the first Model Context Protocol (MCP) server that enables seamless interaction between LLMs and open-source EDA toolchains. \name provides a unified interface that allows LLMs to dynamically orchestrate complete open-source RTL-to-GDSII flows through natural language interaction. The framework also implements backend-aware synthesis optimization by creating a closed-loop feedback mechanism between synthesis decisions and actual physical implementation results.

The challenge of \textbf{fixed template workflows} in current RTL-to-GDSII flows is addressed through \name's flexible protocol architecture, which enables dynamic tool sequence composition and adaptive strategy execution. This approach replaces predetermined scripts with intelligent workflow orchestration, allowing the system to adapt tool sequences based on design requirements and intermediate results.

The issue of \textbf{backend-unaware synthesis optimization} is tackled through a closed-loop optimization methodology where LLMs analyze actual post-layout metrics from physical implementation results to iteratively refine synthesis scripts. This approach replaces the traditional reliance on inaccurate Wire Load Models (WLM) with real backend performance data, enabling synthesis decisions that are informed by actual physical implementation constraints and parasitic effects.

The limitation of \textbf{rigid API call patterns} that constrain LLM capabilities and result in suboptimal performance is resolved through MCP's tool calling architecture. This framework provides quantified efficiency gains by enabling intelligent tool selection with interactions throughout the design process, allowing LLMs to maintain context and make informed decisions based on accumulated design knowledge and feedback.

The challenge of \textbf{insufficient dynamic tool selection} is addressed through \name's agent-based architecture, which enables LLMs to function as intelligent orchestration agents. These agents make sequential decisions based on real-time design metrics and intermediate results, dynamically selecting appropriate tools and optimization strategies as the design progresses through the implementation flow.

Our key contributions are summarized as follows:
\begin{enumerate}
    \item  \textbf{Novel MCP-based EDA Architecture}: First Open-Source MCP server for comprehensive RTL-to-GDSII automation enabling natural language control of integrated open-source EDA toolchains. 
    \item \textbf{Backend-Aware Synthesis Optimization}: Closed-loop methodology bridging documented synthesis-implementation gaps using actual post-layout metrics for parameter tuning. 
    \item \textbf{Dynamic LLM-Controlled Design Flow}: Adaptive orchestration leveraging MCP's proven efficiency advantages for selective step execution and iterative refinement. 
    \item \textbf{Comprehensive Evaluation Framework}: We establish a rigorous design benchmark suite comprising hand-selected designs from OpenCores and various open-source repositories, systematically demonstrating 1-30\% critical path improvements and 5-30\% area reduction compared to default synthesis flows across diverse design complexities.

\end{enumerate}

%% file: sections/background.tex
\section{background}

Traditional RTL-to-GDSII workflows, such as OpenLane and OpenRoad~\cite{openroad2019} ~\cite{openlane}, remain fundamentally constrained by predetermined design flows and rigid orchestration methodologies. The diversity of EDA tasks and the specialized nature of hardware description languages present fundamental challenges for automation, which normally require extensive human labor and expertise. Industry research confirms that point-tool solutions create silos where design changes trigger complete flow re-runs, while design space exploration remains constrained by substantial computational costs~\cite{physical_aware2018, geralla2018optimization}.

Another challenge in design space exploration is that classical synthesis optimization faces significant challenges due to its reliance on Wire Load Models (WLMs), which provide statistical approximations rather than actual routing characteristics, leading to poor correlation between synthesis estimates and physical implementation~\cite{physical_aware2018}. While synthesis-to-place-and-route timing correlation achieved 3-4\% accuracy at older nodes, significant gaps emerge at 28nm and below where interconnect parasitics dominate and WLM reliability degrades~\cite{physical_aware_challenges2025}. Backend feedback limitations exacerbate these correlation gaps, with timing violations discovered only after place-and-route, necessitating expensive iteration cycles.

Design space exploration compounds these issues through computational complexity. The exponential growth in optimization combinations makes exhaustive search unfeasible~\cite{hls_dse_survey2019}, while traditional approaches suffer from either fixed optimization patterns with inadequate coverage or vast explorations lacking sufficient guidance~\cite{dse_challenges2008}. The complexity of directive combinations while simultaneously balancing timing, area, and power constraints remains a fundamental challenge~\cite{dse_optimization2019}.

Current LLMs show promise for design space exploration due to their strong generation capabilities and rule understanding. However, existing integration approaches face several key limitations. First, they rely on fixed API calling sequences and predetermined synthesis flows, preventing dynamic tool selection or adaptive execution strategies~\cite{llm4eda2024, edasurvey}. This constrains LLMs within predetermined boundaries, limiting intelligent design space exploration and conditional tool execution capabilities.
Second, performance issues arise from excessive generation times due to token limits and restart requirements~\cite{verilogeval2023, chipchat2023}. Third, technical constraints include limited hardware-related knowledge in pretrained models and difficulties with specialized HDL syntax, necessitating extensive domain-specific fine-tuning~\cite{chipnemo2023}. These fixed integration patterns prevent the dynamic, adaptive approaches needed for effective design space exploration~\cite{verigen2023, thakur2022benchmarking}.

Model Context Protocol addresses integration challenges by providing standardized interfaces for AI-tool communication, though academic evaluation of its effectiveness in EDA contexts remains limited. MCP enables standardized connections between AI models and external data sources through a client-server architecture that maintains stateful sessions~\cite{mcp_specification2025}. The protocol facilitates dynamic tool discovery and reduces integration complexity from M×N custom connectors to M+N standardized implementations. Standardized messaging protocols enable seamless interaction across diverse systems while breaking down data silos~\cite{mcp_landscape2025}. Security frameworks incorporate authentication mechanisms for remote server connections and flexible resource management through server-initiated completion requests. However, the protocol's effectiveness for complex EDA workflows, token efficiency improvements, and scalability benefits require empirical validation through controlled studies. Current academic literature lacks comprehensive performance comparisons between MCP and traditional API approaches in EDA-specific contexts, representing an important area for future research.

%% file: sections/methodology.tex
\section{Methodology}

\begin{figure}[!ht]
\centering
\includegraphics[width=0.95\linewidth]{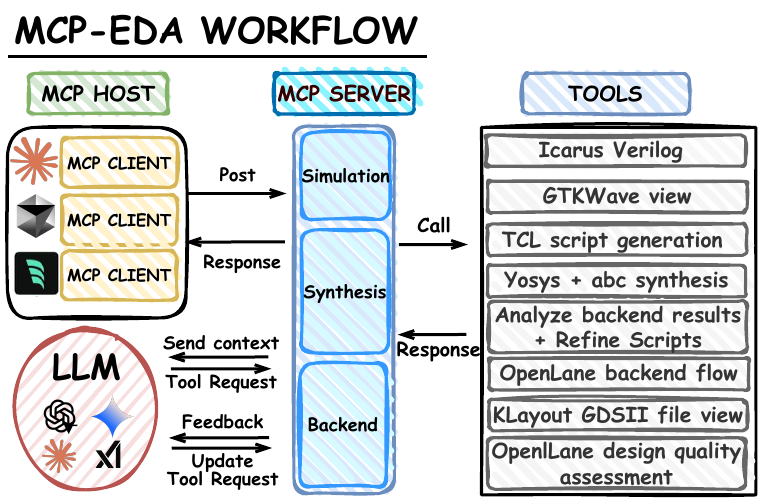}
\caption{\name workflow architecture integrating Model Context Protocol components with EDA tools for backend-aware synthesis optimization. The \textbf{MCP Host} (Claude Desktop/Cursor IDE) provides natural language interface. Multiple \textbf{MCP Clients} manage individual tool connections with stateful sessions. The \textbf{MCP Server} orchestrates tool execution across three domains: \textit{Simulation} (Icarus Verilog, GTKWave), \textit{Synthesis} (TCL generation, Yosys~\cite{wolf2013yosys}+abc~\cite{BraytonMishchenko2010abc}, backend analysis), and \textit{Backend} (OpenLane flow, layout visualization, quality assessment). The \textbf{LLM} receives real backend metrics through feedback loops, enabling iterative script refinement based on actual post-layout performance.}
\label{fig:mcp_workflow}
\end{figure}

\begin{figure}[!ht]
\centering
\includegraphics[width=0.95\linewidth]{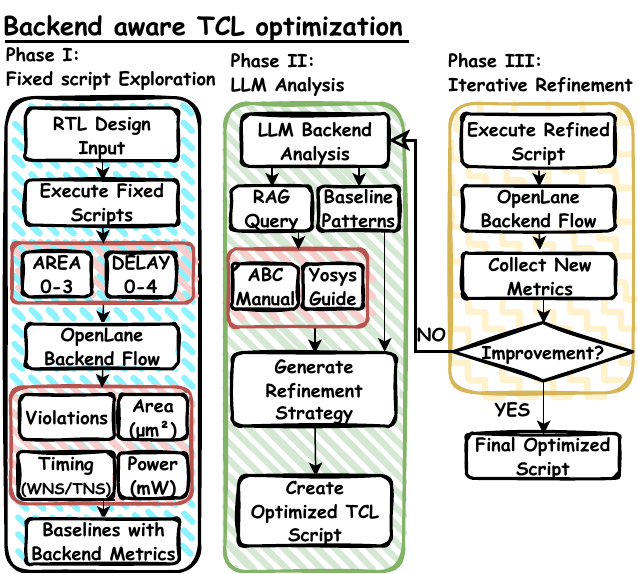}
\caption{Backend-aware TCL optimization workflow showing three-phase iterative synthesis optimization. \textbf{Phase I} establishes baseline metrics through fixed script exploration with OpenLane backend execution. \textbf{Phase II} employs LLM analysis of backend results using RAG query integration with ABC manual and Yosys guides to generate optimized TCL scripts. \textbf{Phase III} implements iterative refinement cycles, executing refined scripts through OpenLane, collecting performance metrics, and continuing optimization until convergence.}
\label{fig:backend_tcl_optimization}
\end{figure}
\subsection{Solution to Challenge 1: Dynamic Workflow Architecture}

\name implements a flexible, adaptive workflow architecture replacing rigid predetermined sequences with intelligent, context-aware design flow orchestration. The \textbf{MCP Host} serves as the primary user interface through Claude Desktop or Cursor IDE environments, providing natural language interaction capabilities. Users specify design objectives, constraints, and optimization targets through conversational prompts, eliminating manual tool configuration.

The host maintains session context across multiple design iterations, enabling complex multi-step optimization workflows that adapt based on intermediate results. Unlike template-based approaches following predetermined sequences, the \name architecture enables dynamic strategy execution where the LLM modifies workflow sequences, skips unnecessary steps, or introduces additional optimization stages based on real-time assessment of design progress and constraints.

\begin{figure}[!t]
\centering
\begin{tcolorbox}[
    colback=promptcolor,
    colframe=borderblue,
    width=\linewidth,
    arc=2mm,
    boxrule=1.5pt,
    title={\textbf{LLM Prompt}},
    fonttitle=\bfseries\large,
    coltitle=white,
    colbacktitle=borderblue,
    fontupper=\sffamily
]
\textbf{Optimization Goal:} Based on ABC synthesis documentation, and the  ABC  command with physical results,  focus on TIMING optimization.
\begin{lstlisting}
Iteration 2: 3971.31 $\mu$m$^2$ area, 0.88ns critical path, 0.000578 $\mu$W power [TIMING IMPROVED: 0.030ns].

+read_constr,$\$${sdc_file};strash;dch;map -B 0.85;buffer -c -N 4; upsize -c;map -B 0.95;buffer -c ... -m
\end{lstlisting}

\end{tcolorbox}

\begin{tcolorbox}[
    colback=responsecolor,
    colframe=bordergreen,
    width=\linewidth,
    arc=2mm,
    boxrule=1.5pt,
    title={\textbf{LLM Response}},
    fonttitle=\bfseries\large,
    coltitle=white,
    colbacktitle=bordergreen,
    fontupper=\sffamily
]
\textbf{Recommended Optimization Strategy:}

Applying balanced buffer mapping with enhanced resynthesis sequence:
\begin{lstlisting}
+read_constr,$\$${sdc_file};strash;$\$${abc_rf};dch;map -B 0.87;buffer -c -N ... -m
\end{lstlisting}
\end{tcolorbox}

\begin{tcolorbox}[
    colback=resultcolor,
    colframe=borderorange,
    width=\linewidth,
    arc=2mm,
    boxrule=1.5pt,
    title={\textbf{Execution Result}},
    fonttitle=\bfseries\large,
    coltitle=white,
    colbacktitle=borderorange,
    fontupper=\sffamily
]
\begin{lstlisting}
Iteration 3: 3913.75 $\mu$m$^2$ area, 0.86ns critical path,  0.000538 $\mu$W power
\end{lstlisting}
\textbf{Summary:} The LLM-guided optimization achieved timing improvement of \texttt{0.050ns} and area reduction of \texttt{1.4\%}.
\end{tcolorbox}
\caption{Example LLM-Guided Synthesis Optimization with Iterative Refinement}
\label{lst:llm_example}
\end{figure}

\subsection{Backend-Aware Synthesis Optimization}

\name's core innovation lies in backend-aware synthesis optimization methodology, creating closed-loop feedback between synthesis decisions and actual physical implementation results. This differs from traditional synthesis flows relying on Wire Load Models or statistical approximations. The complete optimization workflow (Figure~\ref{fig:backend_tcl_optimization}) demonstrates the three-phase iterative process for optimal synthesis configurations.

\subsubsection{Phase I: Fixed Script Exploration and Baseline Establishment}

The optimization process begins with systematic baseline establishment through fixed script exploration. The system executes predefined synthesis scripts with varying optimization levels (AREA 0-3, DELAY 0-4) to establish performance baselines across the design space. Each configuration undergoes complete OpenLane backend flow execution, generating comprehensive metrics including area utilization ($\mu$m²), power consumption (mW), timing performance (WNS/TNS), and constraint violations.

This baseline exploration: (1) establishes performance boundaries and feasible optimization ranges, (2) identifies promising configuration patterns demonstrating effective optimization trade-offs, and (3) provides ground-truth backend metrics enabling accurate correlation between synthesis parameters and physical implementation results.

\subsubsection{Phase II: LLM-Guided Analysis and Refinement Strategy Generation}

Phase II leverages LLM capabilities for intelligent analysis of backend results and strategic refinement generation. The LLM analyzes backend results for timing violations, area inefficiencies, power issues, and routing congestion.

The analysis incorporates domain-specific knowledge through the integration of Retrieval-Augmented Generation(RAG) query with the ABC optimization manual and Yosys synthesis guides, enabling access to specialized synthesis knowledge and proven optimization strategies. This correlates the observed performance with documented techniques to generate refinement strategies for identified limitations.

The LLM generates optimized TCL scripts by analyzing baseline patterns and applying targeted optimization strategies. Unlike fixed template approaches, the LLM dynamically adjusts optimization sequences, selects relevant synthesis algorithms, and configures tool-specific parameters based on specific performance characteristics observed in baseline exploration. 

\subsubsection{Phase III: Iterative Refinement and Convergence}

The final phase implements iterative refinement cycles where optimized TCL scripts undergo execution through OpenLane backend flow, generating new performance metrics for comparison with baseline results. The system evaluates improvements in timing, area, power, and design rule compliance.

When improvements are achieved, the system updates the optimization baseline and initiates subsequent iterations, incorporating lessons from previous attempts. The LLM maintains history to avoid revisiting unsuccessful configurations while identifying promising optimization directions for exploration. Refinement continues until design objectives are satisfied.
\subsubsection{LLM-Guided TCL Script Generation}

The synthesis optimization process begins with LLM-driven TCL script generation, where the language model analyzes RTL design characteristics and user-specified constraints to generate optimized synthesis configurations. The LLM considers design complexity, target frequency requirements, area constraints, and power objectives to select appropriate synthesis strategies.

The script generation process incorporates domain-specific knowledge about synthesis tool capabilities, optimization trade-offs, and common design patterns. The LLM maintains awareness of Yosys synthesis options, abc optimization algorithms, and technology mapping strategies, enabling intelligent selection of synthesis flows tailored to specific design requirements. Figure~\ref{lst:llm_example} demonstrates this iterative optimization approach, where the LLM analyzes baseline performance metrics and generates refined ABC command sequences.

\subsubsection{Backend Metrics Analysis and Feedback}

Following synthesis and place-and-route execution through OpenLane, the system extracts comprehensive backend metrics including actual timing analysis results, real area utilization, power consumption estimates, and routing congestion data. These metrics represent ground-truth implementation characteristics rather than synthesis-stage approximations.

The LLM analyzes backend results to identify performance bottlenecks, timing violations, area inefficiencies, and power optimization opportunities. This includes both quantitative metrics and qualitative design analysis. For timing analysis, the LLM examines critical path delays, setup and hold violations, and clock skew characteristics. Area analysis considers standard cell utilization, routing resource consumption, and macro placement efficiency.

This backend-aware approach optimizes synthesis using actual implementation data rather than estimates, resulting in improved correlation between synthesis predictions and final silicon performance.

\subsection{Standardized LLM-EDA Integration}

\name leverages the Model Context Protocol to establish standardized, flexible communication between Large Language Models and EDA toolchains. The architecture implements a three-tier structure consisting of MCP Host, MCP Clients, and MCP Server components that collectively enable dynamic tool orchestration and stateful workflow management.

\textbf{MCP Clients} function as protocol intermediaries, with each client maintaining dedicated connections to specific tool categories within the EDA workflow. The multi-client architecture provides: (1) isolation of tool failures preventing cascade errors across the entire workflow, (2) parallel execution capabilities for independent design steps, and (3) modular expansion allowing new tools to be integrated without modifying existing client connections.

Each client maintains state and context throughout optimization cycles. This stateful approach enables the LLM to maintain awareness of previous optimization attempts, intermediate results, and design history, facilitating intelligent decision-making that builds upon previous iterations.

\subsection{Intelligent Tool Selection Framework}

The \textbf{MCP Server} orchestrates intelligent tool execution through domain-based organization enabling selective, context-aware tool invocation. The server manages three primary functional domains: The \textit{Simulation} domain integrates Icarus Verilog for RTL simulation and GTKWave for waveform visualization. The \textit{Synthesis} domain encompasses TCL script generation, Yosys synthesis with abc optimization, and backend result analysis capabilities. The \textit{Backend} domain manages OpenLane place-and-route flows, KLayout GDSII visualization, and comprehensive design quality assessment tools.

This domain-based organization enables the LLM to function as an intelligent design space exploration agent, selectively invoking relevant toolsets based on current optimization objectives rather than following predetermined sequences. The LLM dynamically selects tools, execution order, and parameters based on design state and objectives.

The framework tracks optimization history to avoid failed configurations and identify promising directions, transforming rigid EDA workflows into adaptive optimization processes.

%% file: sections/experiments.tex
\begin{figure*}[!t]
    \centering
    \includegraphics[width=0.85\textwidth]{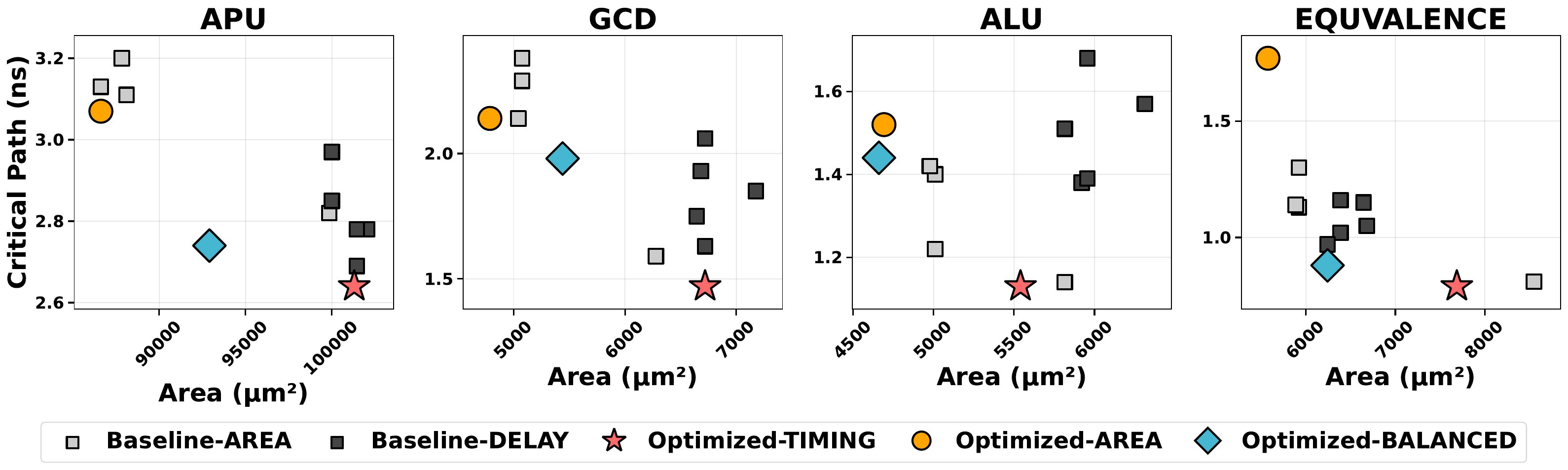}
    \caption{Time vs Area (left) and APU Best Overall Optimization (right)}
    \label{fig:combined}
\end{figure*}

\begin{table*}[!t]
\centering
\caption{Design Performance Comparison: Timing and Area Results}
\renewcommand{\arraystretch}{1.2} 
\footnotesize  
\begin{tabular}{|>{\raggedright\arraybackslash}p{2.5cm}|>{\centering\arraybackslash}p{1.5cm}|>{\centering\arraybackslash}p{1.5cm}|>{\centering\arraybackslash}p{1.5cm}|>{\centering\arraybackslash}p{1.5cm}|>{\centering\arraybackslash}p{1.5cm}|>{\centering\arraybackslash}p{1.5cm}|}
\hline
\textbf{Design} & \textbf{Base} & \textbf{Fixed API} & \textbf{MCP EDA} & \textbf{Base} & \textbf{Fixed API} & \textbf{MCP EDA} \\\hline
& \textbf{Delay (ns)} & \textbf{Delay (ns)} & \textbf{Delay (ns)} & \textbf{Area ($\mu$m²)} & \textbf{Area ($\mu$m²)} & \textbf{Area ($\mu$m²)} \\
\hline
equivalence\_resolver & 0.81 & 0.74 & 0.74 & 5885.64 & 5990.75 & 5574.10 \\
\hline
rs232 & 0.91 & 0.90 & 0.90 & 2781.42 & 2781.42 & 2781.42 \\
\hline
gcd & 2.29 & 2.00 & 1.47 & 5074.87 & 5009.80 & 4785.84 \\
\hline
APU & 2.69 & 2.64 & 2.64 & 86618.07 & 86618.07 & 86618.07 \\
\hline
gcd & 1.59 & 2.00 & 1.47 & 5042.34 & 5009.80 & 4785.84 \\
\hline
UART\_RTO & 0.54 & 0.76 & 0.52 & 1726.66 & 1726.66 & 1707.89 \\
\hline
P16C5x\_ALU & 1.14 & 1.24 & 1.13 & 4977.27 & 4977.27 & 4660.72 \\
\hline
FSM & 0.33 & 0.33 & 0.33 & 52550.40 & 53660.78 & 42040.32 \\
\hline
divide & 2.04 & 2.04 & 2.04 & 18830.56 & 20234.41 & 17902.17 \\\hline
 Log2pipelined& 0.55& 0.55& 0.55& 5405.184& 4057.64&3336.95\\
\specialrule{1pt}{0pt}{0pt}
\textbf{GeoMean} & \textbf{1.05}& \textbf{1.09}& \cellcolor{yellow}\textbf{0.97}& \textbf{8360.21}& \textbf{8436.75}& \cellcolor{yellow}\textbf{7566.24}\\
\hline
\textbf{Ratio} & \textbf{1}& \textbf{1.04}& \cellcolor{yellow}\textbf{0.94}& \textbf{1}& \textbf{1.01}& \cellcolor{yellow}\textbf{0.91}\\
\hline
\end{tabular}
\label{tab:design_timing_area_comparison}
\end{table*}

\section{Experimental Evaluation}

The experimental evaluation of \name's backend-aware synthesis optimization methodology was conducted using a comprehensive suite of digital designs to assess performance improvements across multiple optimization objectives. The evaluation employed OpenLane v2.0 as the baseline implementation flow, comparing default synthesis configurations against \name's LLM-guided optimization strategies.

\subsection{Design Benchmark Suite}

The experimental evaluation utilized a diverse collection of RTL designs spanning different complexity levels and application domains to ensure comprehensive assessment of the optimization methodology. The benchmark suite consists of nine distinct designs sourced primarily from OpenCores~\cite{opencores}, representing various digital signal processing, arithmetic, and communication applications. All designs were synthesized using the SkyWater Sky130 PDK 130nm technology library to ensure consistency in physical implementation characteristics.

Three optimization approaches were evaluated to assess \name's backend-aware synthesis effectiveness:

\textbf{Baseline Configuration}: To establish the effectiveness and targeting effect of different optimization goals, OpenLane's nine existing synthesis options (DELAY 0-4 and AREA 0-3) were executed on all benchmark designs. The best results in Delay and Area are selected to represent ground-truth performance metrics for comparison.

\textbf{Fixed API Template}: Synthesis optimization was performed using fixed API calls with the Claude 4 Sonnet model, utilizing existing physical metrics from the baseline configuration without further feedback. This configuration represents traditional template-based LLM integration approaches that lack iterative backend-aware optimization capabilities.

\textbf{\name Backend-Aware}: Physically-aware synthesis optimization was performed using multiple self-iterations with the Claude 4 Sonnet model, incorporating real backend metrics through the three-phase optimization workflow described in the methodology section. This configuration represents the full \name framework with closed-loop feedback between synthesis decisions and physical implementation results.

All designs were synthesized using Yosys synthesis tool, and the generated netlists were subsequently processed through OpenLane backend tools for place-and-route implementation, ensuring consistent physical implementation flow across all experimental configurations.

\subsection{Experimental Results and Analysis}

As shown in Figure \ref{fig:combined}, four designs are shown with baseline and optimized synthesis to backend results approaches measured and plotted respect of critical path and area, the optimized results have reached the alignment between the requirement (AREA, TIMING, BALANCED) and have also reached the best performance in the required optimization goal while keeping the other metric as uncompromised as possible.

The experimental results in Table~\ref{tab:design_timing_area_comparison} demonstrate the effectiveness of \name's backend-aware synthesis optimization methodology across the diverse benchmark suite. \name consistently achieves superior performance compared to both baseline and fixed API template approaches, with geometric mean results revealing a 6\% improvement in timing performance (0.94 ratio) and a 9\% area utilization (0.91 ratio) compared to the baseline configuration. In contrast, the fixed API template approach shows degraded performance with a 4\% increase in timing delay and 1\% increase in area, highlighting the  importance of backend-aware feedback in synthesis optimization.

\begin{figure}[!t]
    \centering
    \includegraphics[width=0.52 \textwidth]{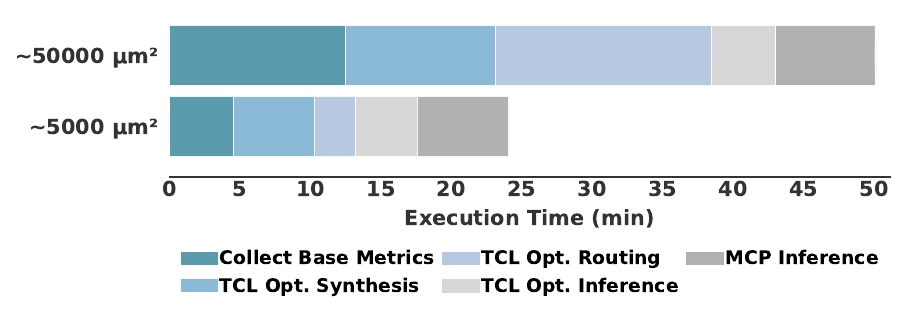}
    \caption{Design Runtime Evaluation}
    \label{fig:runtime}
\end{figure}

As illustrated in Figure~\ref{fig:runtime}, the execution time of the \name workflow exhibits significant variation across designs of different complexities. The backend-aware synthesis optimization employs an iterative refinement process, where each iteration requires complete synthesis and place-and-route execution. For smaller designs with approximately 5,000~$\mu$m² area, each synthesis and routing iteration requires approximately one minute of computation time. For larger designs with areas around 50,000~$\mu$m², this duration reaches approximately three minutes per iteration. The MCP inference overhead remains consistent at approximately seven minutes per design, independent of design complexity.

The experimental results demonstrate that despite computational overhead from iterative optimization cycles, runtime scaling remains tractable for practical design configurations. The primary overhead stems from physical implementation phases—an intrinsic requirement of backend-aware optimization that reflects necessary computational complexity for accurate physical analysis. The strong alignment between user requirements and achieved physical metrics validates \name's effectiveness.

%% file: sections/Limitations.tex
\section{Limitations}
While \name has demonstrated promising timing and area improvements on nine open-source benchmarks under the SkyWater Sky130 PDK, its scope remains limited to small-to-medium RTL designs. Scaling to larger industrial blocks will require addressing longer reports, larger scripts, and increased tool runtime.  
The closed-loop workflow incurs non-negligible compute and runtime overhead: each iteration requires a full synthesis–place-and-route run plus LLM processing, and convergence typically occurs in three to five iterations.  
Dynamic and leakage power metrics were not systematically measured, so the impact of backend-aware scripts on power consumption remains unquantified.  
The framework relies on an off-the-shelf general-purpose LLM (Claude 4) with RAG over Yosys/ABC manuals; domain-specialized models or lightweight fine-tuning could improve EDA-specific reasoning and script quality.  
Finally, \name currently orchestrates only open-source tools and has not been evaluated with commercial synthesis or place-and-route toolchains.

%% file: sections/Future_Work.tex


%% file: sections/conclusion.tex
\section{Conclusion}

\name presents the first Model Context Protocol server that demonstrates LLM-controlled end-to-end RTL-to-GDSII automation with backend-aware synthesis optimization. The system achieves significant performance improvements in critical paths and area reductions compared to baseline synthesis flows through its innovative closed-loop feedback methodology that leverages real post-layout metrics rather than statistical approximations. By integrating a series of open-source EDA tools into a unified LLM-accessible interface, \name transforms rigid template-based EDA workflows into adaptive, intelligent design space exploration processes. The experimental evaluation across nine diverse RTL designs demonstrates the framework's effectiveness in bridging the traditional gap between synthesis estimates and physical implementation reality. Future work will focus on scaling to larger industrial designs through hierarchical optimization and extending support to closed-source commercial EDA tools to demonstrate broader applicability across industry-standard workflows.